\documentclass[runningheads]{llncs}
\usepackage{graphicx,amssymb,amsmath,subfigure}
\usepackage{wrapfig}
\newcommand{\np}{$\mathsf{ NP }$}

\usepackage{hyperref}
\usepackage{xcolor}

\usepackage[T1]{fontenc}

\newcommand{\ptas}{$\mathsf{PTAS}$}

\newcommand{\psat}{\textit{RP3SAT}}

\newcommand{\colb}[1]{{\color{blue}{#1}}}

\newcommand{\stab}{{\textsc{Stabbing-Subdivision}}}
\newcommand{\ind}{{\textsc{Independent-Subdivision}}}
\newcommand{\dom}{{\textsc{Dominating-Subdivision}}}

\newtheorem{theorem1}{Theorem}
\newtheorem{lemma1}{Lemma}

\usepackage[framemethod=tikz]{mdframed}
\newmdenv[backgroundcolor=blue!15,
topline=false,
bottomline=false,
rightline=false,
skipabove=\topsep,
skipbelow=\topsep
]{siderules}

\usepackage{pifont}
\usepackage{lineno}

\begin{document}

\title{Covering and Packing of Rectilinear Subdivision}
\author{Satyabrata Jana\inst{1} \and
Supantha Pandit\inst{2,}\thanks{The author is partially supported by the Indo-US Science \& Technology Forum (IUSSTF) under the SERB Indo-US Postdoctoral Fellowship scheme with grant number 2017/94, Department of Science and Technology, Government of India.}}
\authorrunning{ S Jana and S Pandit}
\institute{Indian Statistical Institute, Kolkata, India \and
State University of New York at Stony Brook, New York, USA\\
\email{\{satyamtma,pantha.pandit\}@gmail.com}}

\maketitle

\begin{abstract}
We study a class of geometric covering and packing problems for bounded regions on the plane. We are given a set of axis-parallel line segments that induces a planar subdivision with a set of bounded (rectilinear) faces. We are interested in the following problems. 
\begin{description}
\item[(P1) \stab:] Stab all bounded faces by selecting a minimum number of points in the plane.
\item[(P2) \ind:] Select a maximum size collection of pairwise non-intersecting bounded faces.
\item[(P3) \dom:] Select a minimum size collection of faces such that any other face has a non-empty intersection (i.e., sharing an edge or a vertex) with some selected faces. 
\end{description} 
We show that these problems are \np-hard. We even prove that these problems are \np-hard when we concentrate only on the rectangular faces of the subdivision. Further, we provide constant factor approximation algorithms for the \stab~problem.

\end{abstract}

\section{Introduction}

Set Cover and Independent Set problems are two well-studied problems in across fields. In Set Cover problem, we are given a set of points and a set of objects, the goal is to find a minimum collection of objects which covers all the points. In Independent Set problems we are given a set of objects and find a maximum collection of a pairwise non-intersecting set of objects. A Dominating Set problem is a variation of the Set Cover problem. In this problem, we are given a set of objects and find a minimum collection of objects such that any remaining object has a non-empty intersection with some chosen objects.

In this paper, we study a variation of the Set Cover, Independent Set, and Dominating Set problems. We are given $m$ axis-parallel line segments that induce a planar subdivision $\mathcal{P}$ with a set $F$ of $n$ bounded rectilinear faces. We formally define these problems as follows.

\begin{siderules}
{\bf (P1) \stab:} We are given a planar subdivision of $n$ bounded faces $F$, find a minimum size points in the plane such that each face in $F$ is stabbed.
\end{siderules}

\vspace{-.3cm}

\begin{siderules}
{\bf (P2) \ind:} We are given a planar subdivision of $n$ bounded faces $F$, find a maximum size objects $F'\subseteq F$ such that any pair of objects in $F'$ is non-intersecting.
\end{siderules}
\vspace{-.3cm}

\begin{siderules}
{\bf (P3) \dom:} We are given a planar subdivision of $n$ bounded faces $F$, find a minimum size objects $F'\subseteq F$ such that any objects in $F\setminus F'$ has a non-empty intersection with an object in $F'$.
\end{siderules}

A special case of the \stab~problem has an application to the art gallery problem \cite{czyzowicz1994guarding}. Suppose a rectangular art gallery is given. The gallery is subdivided into rectangular rooms. Now the question is ``how
many guards are needed to be stationed in the gallery so as to protect all the rooms?'' This problem is nothing but the \stab~problem where the input faces are all rectangular. We also consider the rectilinear rooms (the original input of the \stab~problem) instead of the rectangular rooms in a planar subdivision and ask the same question as  ``how many guards (also name as representatives) are needed to be stationed in the gallery so as to protect all of the rectilinear rooms?''. 

One thing we need to mention that, in this paper, we sometime use rectangles to interpret rectangular faces of a subdivision.



\subsection{Previous Work}

Set Cover, Independent Set, and Dominating Set problems are \np-hard for simple geometric objects such as disks \cite{Fowler1981}, squares \cite{Fowler1981}, rectangles \cite{Fowler1981}, etc. There is a long line of research of these problems and its various variants and special cases \cite{Hochbaum1985,mustafa2010improved,Mustafa2014,Chan2012,Adamaszek2013,Mudgal2015,Chuzhoy2016,Pandit2017,vanLeeuwen2009}. 

Recently, Korman et al. \cite{korman2018line} studied an interesting variation of the Set Cover problem, the Line-Segment Covering problem. In this problem, they cover all the cells of an arrangement formed by a set of line segments in the plane using a minimum number of line segments. They showed that the problem is NP-hard, even when all segments are axis-aligned. In fact,they also proved that it is \np-hard to cover all rectangular cells of the arrangement by a minimum number of axis-parallel line segments. 
  
In \cite{gaur2002constant}, Gaur et al. studied the rectangle stabbing problem. Here \cite{gaur2002constant} given a set of rectangles, the objective is to stab all rectangles with a minimum number of axis-parallel lines. They provided a 2-approximation for this problem.
  
Czyzowicz et al. \cite{czyzowicz1994guarding} considered the guarding problem in rectangular art galleries. They showed that if a rectangular art gallery divided into $ n $ rectangular rooms, then $ \lceil n/2\rceil$ guards are always sufficient to protect all rooms in that rectangular art gallery. They also extend their result in non-rectangular galleries and 3-dimensional art galleries \cite{czyzowicz1994guarding}.

\subsection{Our Results}
 In this paper, we present the following results.

\begin{itemize}
\item[\ding{229}] We first prove that the \stab~problem is \np-hard when we stab all the rectangular faces of the subdivision. Next, we show that the \stab~problem is \np-hard. Further, we provide a 2.083-approximation and a \ptas~for this problem. (Section \ref{sec-stab})

\item[\ding{229}] We prove that the \ind~problem is \np-hard when we consider only the rectangular faces. Then we prove that the \ind~problem is \np-hard. (Section \ref{sec-ind})

\item[\ding{229}] We prove that the \dom~problem is \np-hard by considering only the rectangular faces. Next, we prove that the \dom~problem is \np-hard. (Section \ref{sec-dom})
\end{itemize}


\section{\stab} \label{sec-stab}

\subsection{\np-hardness} \label{sec-stab-np-hard}
\begin{figure}[htbp]
\begin{center}
{\subfigure[ ]{\includegraphics[scale=.42]{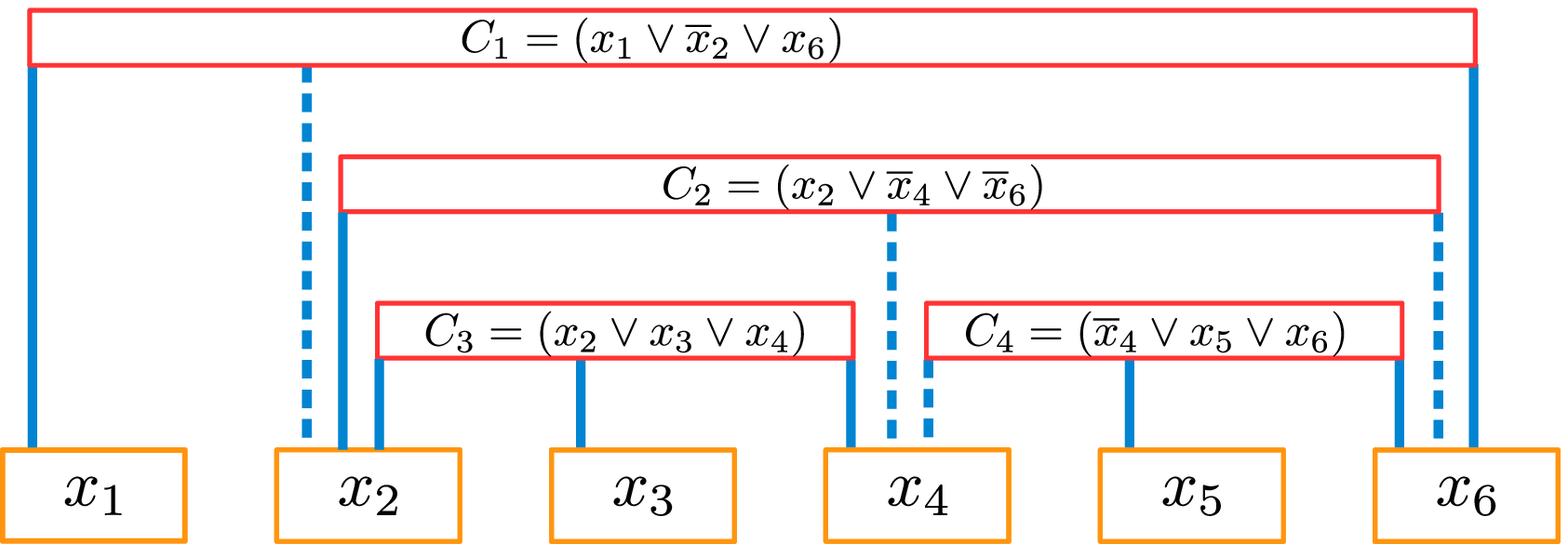}
\label{fig-pnlanar3sat}
}}
\hspace{.15cm}
{\subfigure[ ]{\includegraphics[scale=.55]{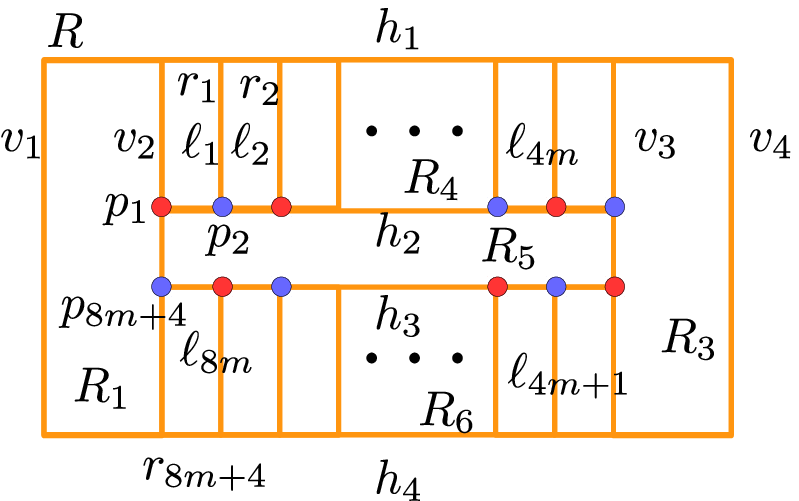}
\label{fig-variable-gadget}
}}
\end{center}
\vspace{-.4cm}
\caption{(a) An instance of the \psat~problem. We only show the clauses which connect to the variables from above.  The solid (resp.\ dotted) lines represent that the variable is positively (resp.\ negatively) present in the corresponding clauses. (b) Structure of a variable gadget. }
\label{fig-stab}
\vspace{-.5cm}
\end{figure}

In this section, we first prove that the \stab~problem is \np-hard when we are restricted to stab only rectangular faces of the subdivision. Next, we modify the construction to show that the \stab~problem is \np-hard.  We give a reduction from the {\it Rectilinear Planar 3SAT (\psat)} Problem. Knuth and Raghunathan \cite{knuth1992problem} proved that this problem is \np-complete. We define this problem as follows. We are given a {\it 3-SAT} formula $\phi$ with $n$ variables $x_1, x_2, \ldots, x_n$ and $m$ clauses $C_1, C_2, \ldots, C_m$ where each clause contains exactly 3 literals. For each variable or clause take a rectangle. The variable rectangles are placed on a horizontal line such that no two of them intersect. The clause rectangles are placed above and below this horizontal line such that they form a nested structure. The clause rectangles connect to the variable rectangles by vertical lines such that no two lines intersect. The objective is to decide whether there is a truth assignment to the variables that satisfies $\phi$. See Figure \ref{fig-pnlanar3sat} for an instance of the \psat~problem. 
We now construct an instance $I$ of \stab~problem from  an instance $\phi$ of the \psat~problem. 
	
\noindent{\bf Variable gadget:}  The gadget of $x_i$ consists of $8m+4$ vertical and $4$ horizontal line segments. See Figure \ref{fig-variable-gadget} for the construction of the gadget. The 4 segments $v_1, v_4, h_1$, and $h_4$ together form  a rectangular region $R$. Next, the 2 vertical segments $v_2$ and $v_3$ partition $R$ vertically into 3 rectangles $R_1$, $R_2$, and $R_3$. Further, two horizontal segments $h_{2}$ and $h_{3}$ partition $R_2$ horizontally into three rectangles $R_4$, $R_5$, and $R_6$. Finally, the $4m$ vertical segments $l_1, l_2, \ldots, l_{4m}$ partition $R_4$ vertically into $4m+1$ small rectangles $r_1,r_2,\ldots,r_{4m+1}$. Similarly, the $4m$ vertical segments $l_{4m+1}, l_{4m+2}, \ldots, l_{8m}$ partition $R_6$ vertically into $4m+1$ small rectangles $r_{4m+2},r_{4m+3},\ldots,r_{8m+2}$. Finally we have the total of $8m+5$ rectangles $R_1, R_3, R_5, r_1,r_2, \ldots, r_{8m+2}$ inside $R$. Clearly, these rectangles except $R_5$ form a cycle of size $8m+4$. Observe that any point along the cycle can stab at most two consecutive regions. Therefore there are two optimal solutions $P^i_1=\{p_1,p_3,\ldots,p_{8m+3}\}$ and $P^i_2=\{p_2,p_4,\ldots,p_{8m+4}\}$ each of size $4m+2$ (Note that these points are not as a part of the input, they are one set of canonical points.). These two solutions are corresponding to the truth value of $x_i$. 
	
	
\noindent {\bf Clause gadget:} The gadget for the clause $C_{\alpha}$ consists of a single rectangle $r_{\alpha}$ that is formed by four line segments. The rectangle $r_{\alpha}$ can be interpreted as the same rectangle as $C_{\alpha}$ in the \psat-problem instance. 
	
\noindent {\bf Interaction:} Now we describe how the clause gadgets interact with the variable gadgets. Observe that the description for the clauses which connect to the variables from above are independent with the clauses which connect to the variables from below. Therefore, we only describe the construction for the clauses which connect to the variables from above.  Let $C^i_1, C^i_2, \ldots, C^i_\tau$ be the left to right order of the clauses which connect to $x_i$ from above. Then we say that $C^i_k$ is the $k$\textsuperscript{th} clause for $x_i$.  For example, $C_3$, $C_2$, and $C_4$  are the 1\textsuperscript{st}, 2\textsuperscript{nd}, and 3\textsuperscript{rd} clause for the variable $x_4$ in Figure \ref{fig-pnlanar3sat}. Let $C_\alpha$ be a clause containing the variable $x_i, x_j,x_t$. We say that the clause $C_\alpha$ is the ${k_1}$, ${k_2}$, and ${k_3}$\textsuperscript{th} clause for variable $x_i$, $x_j$, and $x_t$ respectively based on the above ordering. For example, $C_3$ is the 3\textsuperscript{rd}, 1\textsuperscript{st}, and 1\textsuperscript{st} clause for variable $x_2$, $x_3$, and $x_4$ respectively in Figure \ref{fig-pnlanar3sat}. Let $r_\alpha$ be the rectangle corresponding to $C_\alpha$. Now we have the following cases.


\begin{itemize}
\item[$\bullet$] If $x_i$ appears as a positive literal in clause $C_\alpha$, then extend the 3 segments $l_{4k_1-3}$, $l_{4k_1-2}$, and $l_{4k_1-1}$ vertically upward such that it touches the bottom boundary of the rectangle $r_\alpha$. 

\item[$\bullet$] If $x_i$ appears as a negative literal in clause $C_\alpha$, then extend the 3 segments $l_{4k_1-2}$, $l_{4k_1-1}$, and $l_{4k_1}$ vertically upward such that it touches the bottom boundary of the rectangle $r_\alpha$.
\end{itemize}

	\begin{figure}[ht!]
		\begin{center}
			\includegraphics[scale=.68]{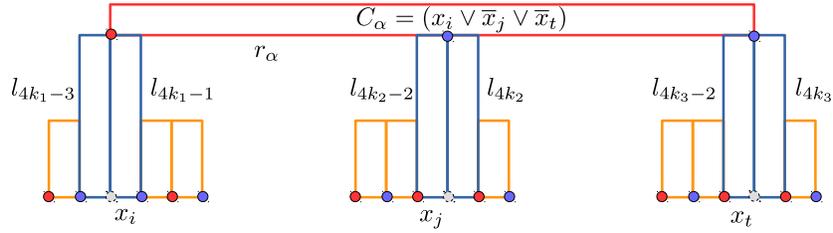}
			\vspace{-.2cm}
			\caption{Variable clause interaction.}
			\label{fig-variable-clause-interact}
		\end{center}
		\vspace{-.6cm}
	\end{figure}

The similar construction can be done for $x_j$ and $x_t$ by replacing $k_1$ with $k_2$ and $ k_3$ respectively. The whole construction is shown in Figure \ref{fig-variable-clause-interact}. Note that, we break the horizontal segment $h_1$ in the variable gadgets into smaller intervals and shifted the intervals vertically along with the extension of the vertical lines. This completes the construction and clearly, it can be done in polynomial time.
	
	
\begin{lemma1}\label{lemma-stab-np-hard} $\phi$ is satisfiable if and only if there is a solution to the \stab~problem to stab only rectangular faces with $n(4m+2)$ points. 
\end{lemma1}
	
\begin{proof} Assume that $\phi$ is satisfiable i.e., we have a truth assignment of the variables in $\phi$. Now consider a variable $x_i$. If $x_i$ is true, we select the set $P^i_1$, otherwise we select the set $P^i_2$. Clearly, the $n(4m+2)$ selected points corresponding to all variable gadgets stab all the rectangular faces of the construction.
	
On the other hand, assume that \stab~problem has a solution with $n(4m+2)$ points. Observe that at least $(4m+2)$ points are needed to stab all the faces of a variable gadget. Since the rectangular faces of variable gadgets are disjoint from each other, exactly $(4m+2)$ points must be selected from each variable gadget. Now there are exactly two solutions of size $(4m+2)$, either $P_1^i$ or $P_2^i$. Therefore, we set variable $x_i$ to be true if $P_1^i$ is selected from the gadget of $x_i$, otherwise we set $x_i$ to be false. Note that for each clause $C_\alpha$ the six faces corresponding to three literals it contains, touches the rectangle $r_\alpha$. Since $r_\alpha$ is stabbed, at least one of the selected points must be chosen in the solution. Such a point is either in one of the sets $P^i_1$ or $P_2^i$ of the corresponding variable gadget based on whether the variable is positively or negatively present in that clause. Hence, the above assignment is a satisfying assignment.
\hfill $\qed$
\end{proof}
	
	
\begin{theorem1}\label{thm-np-hard} The \stab~problem is \np-hard for stabbing only rectangular faces of a subdivision.
\end{theorem1}
	
\noindent {\bf The \stab~problem for stabbing all rectilinear faces:}	We now prove that it is also \np-hard to stab all (rectilinear) faces of a subdivision. We modify the above \np-hardness to prove the hardness. Note that, after embedding the gadgets on the plane, the subdivision creates three types of faces, (i) \colb{variable faces}: the faces interior to a variable gadget (note that all variable faces are rectangular), (ii) \colb{clause faces}: rectangular faces associated with clause gadgets, and (iii) \colb{outer faces}: faces that are not included to any of (i) or (ii).

Note that, in the proof of the Lemma \ref{lemma-stab-np-hard}, we assume that the canonical points (set of $4m+2$ points to stab $8m+5$ rectangles in a variable gadget) are on the lines $h_2$ and $h_3$ (see Figure \ref{fig-variable-gadget}). However, in this case, we keep only one point either on $h_2$ or on $h_3$ (to stab the rectangle $R_5$) and out of the remaining points that are on $h_2$ we shift them vertically upward to $h_1$ and that are on $h_3$ we shift them vertically downward to $h_4$.
Clearly, any outer face includes some variable canonical points. Hence, with this modification, it is immediate that the Lemma \ref{lemma-stab-np-hard} is true even when we are restricted to stab all the faces of the subdivision.

\subsection{Approximation algorithms}\label{sec-stab-approx}

\subsubsection{Factor 2.083 approximation} 
We are given $m$ axis-parallel line segments that induce a planar subdivision $\mathcal{P}$ with a set $F$ of $n$ bounded rectilinear faces. To provide the approximation algorithm, we transform any instance of the \stab~problem into an instance of the Set Cover problem where the size of each set is at most 4. 
Observe that, there exists an optimal solution to the \stab~problem that only contains vertices of $\mathcal{P}$ (we can call them as corner points of $F$). Also, any corner point of $F$ can stab at most $4$ rectilinear faces in $\mathcal{P}$.


We now create an instance of the Set Cover problem as follows. The set of elements is the set of all faces and the collection is all sets of faces corresponding to the corner points of $F$. Note that each set in the collection is of size at most 4, since any corner point can stab at most 4 faces. This Set Cover instance admits a 2.083 ($H_4$ i.e., harmonic series sum of the first 4 terms) factor approximation \cite{Vazirani2001}. Hence we have the following theorem.

\begin{theorem1}
There exists a $2.083$ factor approximation algorithm for \stab~problem in a planar subdivision by rectilinear line segments.
\end{theorem1}

\subsection{PTAS via Local Search Algorithm}
In this section, we show that a local search framework \cite{mustafa2010improved}  leads to a \ptas~for the \stab~problem. We are given a planar subdivision with a set $F$ of $n$ bounded faces. Note that, we can choose points only from the vertex set $V$ of the subdivision. Therefore, ${\cal R}= (V, F)$ be the given range space. Clearly $V$ is a feasible solution to the \stab~problem. 
We apply the  \colb{$k$-level local search} ($k$ is a given parameter) as follows. 
\begin{enumerate}
\item Let  $X$ be some feasible solution to the \stab~problem (initially take $X$ as $V$).
\item Do the following:
		    \begin{enumerate}
		        \item Search for $X' \subseteq X$ and $Y$ such that $ |Y| \subseteq V$, $|Y| < |X'| \leqslant k$ and $(X \setminus X') \cup Y$ is a feasible solution.
               \item If such $X'$ and $Y$ exist, update $X$ with $(X \setminus X') \cup Y$ and repeat the above step. Otherwise, return $X$ and stop.  
		    \end{enumerate}

	\end{enumerate}

It is easy to see that the running time of the algorithm is polynomial. Further, the local search algorithm always returns a local optimum solution.  A feasible solution $X$ is said to be a local optimum if there is no $X'$ exists in Step 2(a) in the above algorithm.
We show that given any $\epsilon >0$, a $ O (1/\epsilon^2)$-level local search returns a hitting set of size at most  $(1+ \epsilon)$ times an optimal hitting set for $\cal R$. \\

\noindent ${\textit{Locality condition}}$
	(\cite{mustafa2010improved}):
	A range space $\cal R$ $ = (V, F)$ satisfies the locality condition if for
	any two disjoint subsets $R, B \subseteq V $, it is possible to construct a planar bipartite graph $G = (R \cup B, E)$ with all edges going between $R$ and $B$ such that for any $f \in  F$, there exist two vertices $u \in f \cap R$ and $v \in f \cap B$  such
	that edge $(u, v) \in E$.
\begin{theorem1}\label{theorem_neighberwood}\cite{mustafa2010improved}
Let $\cal R$ $=(V, F)$ be a range space satisfying the locality condition. Let $R \subseteq  V$ be an optimal hitting set for $ F$, and $B \subseteq  V $ be the hitting set returned by a $k$-level local search. Furthermore, assume $R \cap B = \phi$. Then there exists a planar bipartite graph $G = (R \cup B, E)$ such that for every subset $B'\subseteq B$ of size at most $k$, $|N_{G}(B')| \geq |B'|$ where $N_{G}(W)$ denotes the set of all neighbours of the vertices of $W$ in $G.$
\end{theorem1}
The following lemma implies that given any $\epsilon>0$, a $k$-level local search with $\epsilon = \dfrac{c}{\sqrt{k}}$ gives a $(1 + \epsilon)$-approximation for the \stab~problem.
\begin{lemma1}\label{lemma_on_ptas}\cite{mustafa2010improved} 
Let $G = (R \cup B, E)$ be a bipartite planar graph on red and blue vertex sets $R$ and $B$, $|R| \geq 2$, such that for every subset $B' \subseteq B$ of size at most $k$, where $k$ is a large enough number, $|N_{G}(B')| \geq |B'|$. Then $|B| \leq (1 + \dfrac{c}{\sqrt{k}}) |R|$, where $c$ is a constant.
\end{lemma1}
	
\noindent {\bf \ptas~for the \stab~problem:} Let $R$ (red) and $B$ (blue) be disjoint subsets of the vertices in planar subdivision $\mathcal{P}$ where $R$ and $B$ be an optimum solution and the solution returned by the $k$-level local search respectively. For simplicity, we assume that $R \cap B= \phi $. Otherwise, we can remove the common elements from each of $R$ and $B$, and then do the similar analysis. As we remove the same number of elements from both $R$ and $B$, the approximation ratio of the original instance is at most the approximation ratio of the restricted one. We construct the required graph $G$ on the vertices $R \cup B$ in the following way. Since $R$ and $B$ are  feasible solutions of the \stab~problem,  every face $f \in F$  must contain at least one red and one blue point. We simply join exactly one pair of red and blue points by an edge for each face $f \in F$. Clearly, the edge for a face $f \in F$ lies completely inside $f$. Therefore $G$ becomes a planar bipartite graph and hence $\cal R$ satisfies the locality condition. Therefore, from Theorem \ref{theorem_neighberwood} and Lemma \ref{lemma_on_ptas},  we say that the \stab~problem admits a \ptas.

\section{\ind}\label{sec-ind}
	
In this section, we prove that the \ind~problem is \np-hard by giving a reduction from the \psat~problem. The reduction follows the same line of the reduction presented in Section \ref{sec-stab}. We construct an instance $I$ of the \ind~problem from an instance $\phi$ of the \psat~problem and prove that the construction is correct. 
	
\begin{figure}[ht!]
		\begin{center}	\includegraphics[scale=.75]{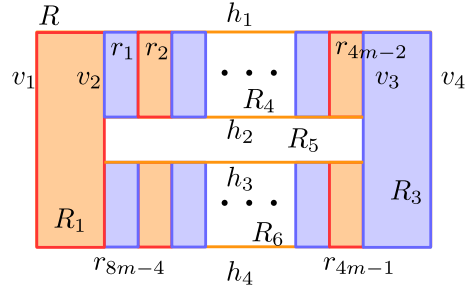}
		\vspace{-.0cm}
			\caption{Structure of a variable gadget.}
			\label{fig-variable-gadget-ind-set}
		\end{center}
		\vspace{-.5cm}
	\end{figure}

	
\noindent {\bf Variable gadget:}  The variable gadget is similar to the variable gadget that is described in the Section \ref{sec-stab}. See Figure \ref{fig-variable-gadget-ind-set} for the construction of a variable gadget. The difference of this variable gadget from the gadget in the Section \ref{sec-stab} is that we partition $R_4$ into $4m-2$ smaller rectangles $r_1,r_2,\ldots,r_{4m-2}$ and  $R_6$ into $4m-2$ smaller rectangles $r_{4m-1},r_{4m},\ldots,r_{8m-4}$. Finally, we have the total of $8m-1$ rectangles $R_1, R_3, R_5, r_1,r_2, \ldots, r_{8m-4}$ inside $R$. Notice that, these rectangles except $R_5$ form a cycle of size $8m-2$. Therefore there are exactly two optimal solutions $S^i_1=\{R_3, r_1,r_3,\ldots,r_{4m-3}, r_{4m},r_{4m+2},\ldots,r_{8m-4}\}$ and $S^i_2=\{R_1, r_2,r_4,\ldots,r_{4m-2}, r_{4m-1}, r_{4m+1},\ldots, r_{8m-5}\}$, each with size $4m-1$. These two solutions are corresponding to the truth values of the variable $x_i$. 

	\vspace{-.4cm}
	
\begin{figure}[ht!]
		\begin{center}
			\includegraphics[scale=.6]{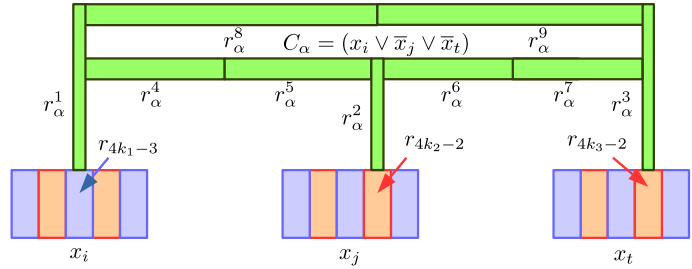}
		\vspace{-.2cm}	\caption{Variable clause interaction.}
			\label{fig-variable-clause-interact-ind-set}
		\end{center}
		\vspace{-.5cm}
	\end{figure}

\noindent {\bf Clause gadget:} The gadget of the clause $C_\alpha$ includes 9 rectangles $r_\alpha^1, r_\alpha^2, \ldots, r_\alpha^9$ (see green rectangles in Figure \ref{fig-variable-clause-interact-ind-set}. The six rectangles $r_\alpha^4, r_\alpha^5, \ldots, r_\alpha^9$ are placed inside the rectangle of $C_\alpha$ in the \psat-problem instance and the other three rectangles $r_\alpha^1, r_\alpha^2, r_\alpha^3$ are corresponding to the three vertical legs between $C_\alpha$ and the three variables it contains. Note that there is another rectangle present in the clause gadget bounded by the above 9 rectangles. However, this rectangle has no effect in the reduction, since picking this rectangle makes other 9 rectangles invalid (can not be selected).
	
\noindent {\bf Interaction:} Here also we describe the construction for the clauses that connect to the variables from above, since the construction is similar and independent from the clauses that connect to the variables from below. Let $C_\alpha$ be a clause containing the variables $x_i, x_j,x_t$. Also assume that this is the left to right order of these variables in which they appear in $\phi$. Using the similar way as before (Section \ref{sec-stab}), we say that the clause $C_\alpha$ is the ${k_1}$, ${k_2}$, and ${k_3}$\textsuperscript{th} clause for the variables $x_i$, $x_j$, and $x_t$ respectively. 
	
	
	
\begin{itemize}
\item[$\bullet$] If $x_i$ appears as a positive literal in the clause $C_\alpha$, then attach the rectangle $r_\alpha^1$ to the rectangle $r_{4k_1-3}$. 
\item[$\bullet$] If $x_i$ appears as a negative literal in clause $C_\alpha$, then attach the rectangle  $r_\alpha^1$ to the rectangle $r_{4k_1-2}$. 
\end{itemize}

The similar construction can be done for $x_j$ by replacing $r_\alpha^1$ and $k_1$ with $r_\alpha^2$ and $k_2$ respectively and for $x_t$ by replacing $r_\alpha^1$ and $k_1$ with $r_\alpha^3$  and $k_3$ respectively. The whole construction is depicted in Figure \ref{fig-variable-clause-interact-ind-set}. Clearly, the construction can be done in polynomial time. We now prove the correctness of the construction.

\begin{lemma1}\label{lemma-ind-np-hard} $\phi$ is satisfiable if and only if there is a solution of size $n(4m-1) +4m$ to \ind~problem while considering only rectangular faces.
\end{lemma1}
\begin{proof} Assume that $\phi$ has a satisfying assignment. For the variable $x_{i}$, if $x_{i}$ is true, select the set $S^i_{2}$, otherwise select the set $S^i_{1}$. Since each set is of cardinality $(4m-1)$, clearly we select $n(4m-1)$ independent rectangles across all variable gadgets. Now let $C_\alpha$ be a clause containing variables $x_i, x_j,x_t$. Since $C_\alpha$ is satisfiable at least one of the three rectangles $r_\alpha^1, r_\alpha^2, r_\alpha^3$ is free to choose in a solution. This implies we can select exactly 4 rectangles from the gadget of $C_\alpha$. We can picked 4 rectangles independently from each clause gadget. Hence, in total we can select $n(4m-1) +4m$ rectangles.


On the other hand, assume that the \ind~problem has a solution $S$ with $n(4m-1)+4m$ rectangles. Note that for each variable gadget the size of an optimal independent set is $(4m-1)$, either the set $S_1^i$  or $S_2^i$. We set the variable $x_i$ to be true if $S_2^i$ is selected from the gadget of $x_i$, otherwise we set $x_i$ to be false. Now we have to show that this assignment is a satisfying assignment for $\phi$ i.e, each clause of $\phi$ is satisfied. Since the variable gadgets are independent, there are at most $n(4m-1)$ rectangles from the variable gadgets belongs to $S$. Also since the size of the solution is $n(4m-1)+4m$, from each clause gadget exactly 4 rectangles is in $S$. Let $C_\alpha$ be a clause containing variables $x_i, x_j,x_t$. As there are 4 independent rectangles from the set $\{r_\alpha^1, r_\alpha^2, \ldots, r_\alpha^9\}$, so one must be from the set $\{r_\alpha^1, r_\alpha^2, r_\alpha^3\}$ that is in the given solution. W.l.o.g. let $r_\alpha^1$ be present, then surely $x_{i}$ is a true variable as our assignment. Hence the above assignment is a satisfying assignment.
\hfill $\qed$
\end{proof}
	
	
\begin{theorem1}\label{thm-np-hard} The \ind~problem is \np-hard by considering only rectangular faces of a subdivision.
\end{theorem1}

\noindent {\bf The \ind~problem for  all rectilinear faces:} We now prove that it is also \np-hard to find a maximum independent set of rectilinear faces in a subdivision. After embedding the construction on the plane, the subdivision creates three types of faces, (i) \colb{variable faces:} The faces that are interior to a variable gadget, (ii) \colb{clause faces:} the faces associated with the clause gadgets, and (iii) \colb{outer faces:} any other faces that are not included to any of (i) or (ii).

Visualize that we are attaching each clause gadget one by one with the variable gadgets. Then each clause gadget creates two additional rectilinear faces, both sides of the rectangle corresponding to the middle leg. Note that, each such face is adjacent with at least 4 clause rectangles and at least 4 variable rectangles. Therefore, picking one of these new faces to the optimal solution makes the solution size strictly less than the original. Therefore, even if we consider all rectilinear faces, Lemma \ref{lemma-ind-np-hard} holds and so Theorem \ref{thm-np-hard}.


\section{\dom}\label{sec-dom}
In this section, we prove that the \dom~problem is \np-hard. We give a reduction from the \psat~problem similar to Section \ref{sec-ind}. 

We construct an instance $I$ of the \dom~problem from an instance $\phi$ of the \psat~problem and prove that the construction is correct. 

\begin{figure}[ht!]
\begin{center}
\includegraphics[scale=.75]{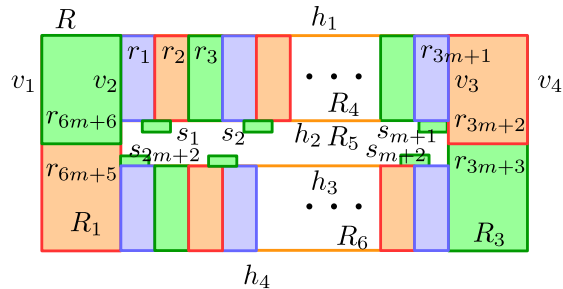}
\vspace{-.0cm}
\caption{Structure of a variable gadget.}
\label{fig-variable-gadget-dom-set}
\end{center}
\vspace{-.4cm}
\end{figure}

\noindent {\bf Variable gadget:}  Variable gadgets are similar to the variable gadgets that is described in Section \ref{sec-stab}. The difference between this variable gadget and that of in Section \ref{sec-stab} is as follows. We partition $R_4$ into $3m+1$ small rectangles $r_1,r_2,\ldots,r_{3m+1}$ and  $R_6$ into $3m+1$ small rectangles $r_{3m+4},r_{3m+5},\ldots,r_{6m+4}$. We partition $R_1$ into two rectangles $r_{6m+6}$, $r_{6m+5}$  and $R_3$ into $r_{3m+2}$, $r_{3m+3}$. Next we take $2m+2$ mutually independent rectangles  $s_1, s_2,\ldots ,s_{2m+2}$ inside $R_5$ such that rectangle $s_i$ touches the two regions $r_{3i-2}$ and $r_{3i-1}$, for $1\leq i\leq 2m+2$. Finally we have a total of $8m+8$ rectangles $r_1, r_2,\ldots,r_{6m+6}, s_1,s_2, \ldots, s_{2m+2}$ inside $R$. Figure \ref{fig-variable-gadget-dom-set} illustrate the construction of a variable gadget just described.

\vspace{-.4cm}

\begin{lemma1}\label{lemma-dom-np-hard}
There exists exactly two optimal dominating sets of rectangles, $D_{1}^i =\{r_1, r_4, \ldots, r_{6m+4}\}$ and
$D_{2}^i = \{r_2, r_5, \ldots, r_{6m+5}\}$, for the gadget of $x_i$.
\end{lemma1}

\begin{proof} There is no rectangle that can dominate more than 4 rectangles. Since there are in total $(8m+8)$ rectangles, any dominating set cannot have size less than $(2m+2)$. Further, both $D_{1}^i$ and $D_{2}^i$, each of size $(2m+2)$, dominate all the faces of the subdivision and hence they are optimal solutions. Now we show that there is no other optimal solution.

Clearly, no rectangle of the form $r_{3k}$ or $s_k$ where $1\leq k \leq (2m+2)$ can be a part of an optimal solution, since each of them dominates exactly 3 rectangles. As a result, any optimal solution contains only rectangles of the form $r_{3k-1}$ or $r_{3k-2}$, for $1\leq k \leq (2m+2)$. Also, two rectangles, one of the form $r_{3k-1}$ and other of the form $r_{3k-2}$, together cannot be a part of any optimal solution.
\hfill $\qed$
\end{proof}

\begin{figure}[ht!]
\vspace{-.1cm}
\begin{center}
\includegraphics[scale=.6]{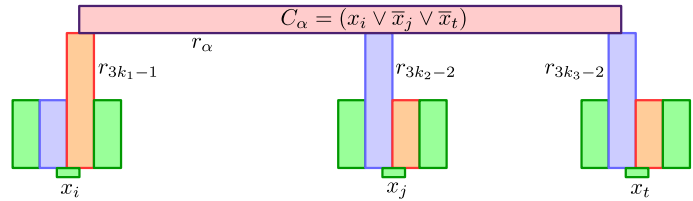}
\vspace{-.2cm}
\caption{Variable clause interaction.}
\label{fig-variable-clause-interact-dom-set}
\end{center}
\vspace{-.7cm}
\end{figure}

\noindent {\bf Clause gadget:} The gadget for the clause $C_\alpha$ is a rectangle $r_\alpha$ (Figure \ref{fig-variable-clause-interact-dom-set}). 

\noindent {\bf Interaction:} Here we describe the construction for the clauses that connect to the variables from above. A similar construction can be done for the clauses that connect to the variables from below. As before, we interpret $C_\alpha$ that contains variables $x_i$, $x_j$, and $x_t$ as the ${k_1}$, ${k_2}$, and ${k_3}$\textsuperscript{th} clause for the variables $x_i$, $x_j$, and $x_t$ respectively.


\begin{itemize}
\item[$\bullet$] If $x_i$ appears as a positive literal in the clause $C_\alpha$, then we extend the rectangle $r_{3k_1-1}$ vertically upward such that it touches the rectangle $r_\alpha$. 
\item[$\bullet$] If $x_i$ appears as a negative literal in the clause $C_\alpha$, then we extend the rectangle $r_{3k_1-2}$ vertically upward such that it touches the rectangle $r_\alpha$. 
\end{itemize}

We make the similar construction for $x_j$ and $x_t$ by replacing $k_1$ with $k_2$ and $k_3$ respectively. The whole construction is depicted in Figure \ref{fig-variable-clause-interact-dom-set}.  Clearly, the construction can be done in polynomial time. We now prove the correctness.

\begin{lemma1}\label{lemma-dom-np-hard-1} $\phi$ is satisfiable if and only if there is a solution of size $n(2m+2)$ to the \dom~problem while considering only rectangular faces.
\end{lemma1}
\begin{proof}
Assume that $\phi$ is satisfiable i.e., we have a truth assignment to the variables of $\phi$. For the variable $x_i$, if $x_i$ is true we select the set $D^i_2$, otherwise we select the set $D^i_1$. Clearly, the $n(2m+2)$ selected rectangles corresponding to all the variable gadgets dominate all the rectangular faces of the subdivision. 

On the other hand, assume that the \dom~problem has a solution with $n(2m+2)$ rectangles. Observe that at least $(2m+2)$ rectangles are needed to dominate all the rectangular faces of a variable gadget. Since the rectangular faces of variable gadgets are disjoint from each other and the size of the solution is $n(2m+2)$, from each variable gadget exactly $(2m+2)$ rectangles must be selected. Therefore, we set variable $x_i$ to be true if $D_2^i$ is selected from the gadget of $x_i$, otherwise we set $x_i$ to be false. Note that for each clause $C_\alpha$ the three rectangles corresponding to the three literals it contains attach to the rectangle $r_\alpha$. Since $r_\alpha$ is dominated, at least one of these three rectangles is chosen in the solution. Such a rectangle is either in $D^i_2$ or $D_1^i$ of the corresponding variable gadget based on whether the variable is positively or negatively present in that clause. Hence, the above assignment is a satisfying assignment.
\hfill $\qed$
\end{proof}


\begin{theorem1}\label{thm-dom-np-hard} The \dom~problem is \np-hard when we are constrained to dominate all the rectangular faces of a subdivision.
\end{theorem1}

\noindent{\bf The \dom~problem for  all rectilinear faces:} We only modify the variable gadgets such that it has exactly two distinct optimal 
 \begin{wrapfigure}{r}{0.48\textwidth}
\vspace{-1.0cm}
\begin{center}
\includegraphics[scale=.6]{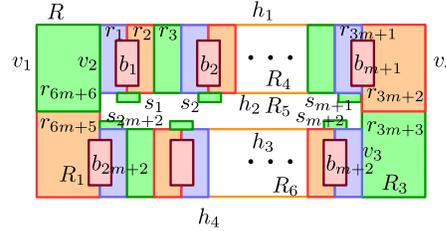}
\vspace{-.3cm}
\caption{Modified variable gadget.}
\vspace{-.4cm}
\label{fig-variable-gadget-modification-dom-set}
\end{center}
\vspace{-.5cm}
\end{wrapfigure}
solutions and the rest of the construction and the proofs remain the same. We take $2m+2$ rectangles $b_1,b_2,\ldots,b_{2m+2}$. 
We place the rectangle $b_i$ in between the rectangles $r_{3i-2}$ and $r_{3i-1}$, for $1\leq i\leq 2m+2$ of the variable gadget shown in Figure \ref{fig-variable-gadget-dom-set} (see Figure \ref{fig-variable-gadget-modification-dom-set}). These additional rectangles enforce not to choose $R_5$ in an optimal solution. Now it is easy to verify that the Lemma \ref{lemma-dom-np-hard} remains true for this modified gadget even when we consider all the bounded faces of the subdivision. 




%

\end{document}